\documentclass[9pt,conference]{IEEEtran}
\IEEEoverridecommandlockouts
\usepackage{cite}
\usepackage{amsmath,amssymb,amsfonts}
\usepackage{algorithmic}
\usepackage{graphicx}
\usepackage{textcomp}
\usepackage{xcolor}
\usepackage{blindtext}
\usepackage{cite}
\usepackage[caption=false]{subfig}
\usepackage{booktabs}
\usepackage{multirow}
\usepackage{hyperref}
\usepackage{pgfplots}
\usepackage{float}
\usepackage{tikz}
\usepackage{multicol}
\pgfplotsset{compat=newest}
\usetikzlibrary{plotmarks, matrix}
\usepgfplotslibrary{external}
\usepgfplotslibrary{groupplots}
\usepgfplotslibrary{colorbrewer}
\usepgfplotslibrary{statistics}
\usepgfplotslibrary{units}
\usepgfplotslibrary{colormaps}
\usepgfplotslibrary{fillbetween}
\usepgfplotslibrary{patchplots}
\usepackage{pifont}
\usepackage{tikz}
\usetikzlibrary{shapes,calc}

\newcommand{\greensub}[1]{%
  \tikz[baseline=(n.base)]{
    \node[draw=green!60!black,           
          fill=green!10,                 
          rounded corners=2pt,           
          inner sep=1pt,                 
          font=\tiny               
         ] (n) {#1};
  }%
}

\newcommand{\redsub}[1]{%
  \tikz[baseline=(n.base)]{
    \node[draw=red!60!black,            
          fill=red!10,                  
          rounded corners=2pt,          
          inner sep=1pt,                
          font=\tiny              
         ] (n) {#1};
  }%
}
\def\BibTeX{{\rm B\kern-.05em{\sc i\kern-.025em b}\kern-.08em
    T\kern-.1667em\lower.7ex\hbox{E}\kern-.125emX}}
\begin{document}

\title{CO-VADA: A Confidence-Oriented Voice Augmentation Debiasing Approach for Fair Speech Emotion Recognition}


\author{\IEEEauthorblockN{
Yun-Shao Tsai\textsuperscript{*}\thanks{\textsuperscript{*}Equal contribution.}}
\IEEEauthorblockA{
\textit{National Taiwan University}\\
Taipei, Taiwan \\
r14942093@ntu.edu.tw}
\and
\IEEEauthorblockN{Yi-Cheng Lin\textsuperscript{*}}
\IEEEauthorblockA{
\textit{National Taiwan University}\\
Taipei, Taiwan \\
f12942075@ntu.edu.tw}
\and
\IEEEauthorblockN{Huang-Cheng Chou}
\IEEEauthorblockA{
\textit{University of Southern California}\\
Los Angeles, California, USA \\
huangchengchou@gmail.com}
\and
\IEEEauthorblockN{Hung-yi Lee}
\IEEEauthorblockA{
\textit{National Taiwan University}\\
Taipei, Taiwan \\
hungyilee@ntu.edu.tw}
}
\maketitle

\begin{abstract}
Bias in speech emotion recognition (SER) systems often stems from spurious correlations between speaker characteristics and emotional labels, leading to unfair predictions across demographic groups.
Many existing debiasing methods require model-specific changes or demographic annotations, limiting their practical use.
We present \textbf{CO-VADA}, a \textit{Confidence-Oriented Voice Augmentation Debiasing Approach} that mitigates bias without modifying model architecture or relying on demographic information.
CO-VADA identifies training samples that reflect bias patterns present in the training data and then applies voice conversion to alter irrelevant attributes and generate samples.
These augmented samples introduce speaker variations that differ from dominant patterns in the data, guiding the model to focus more on emotion-relevant features.
Our framework is compatible with various SER models and voice conversion tools, making it a scalable and practical solution for improving fairness in SER systems.

\end{abstract}


\begin{figure*}[h]
  \centering
  \includegraphics[width=\linewidth]{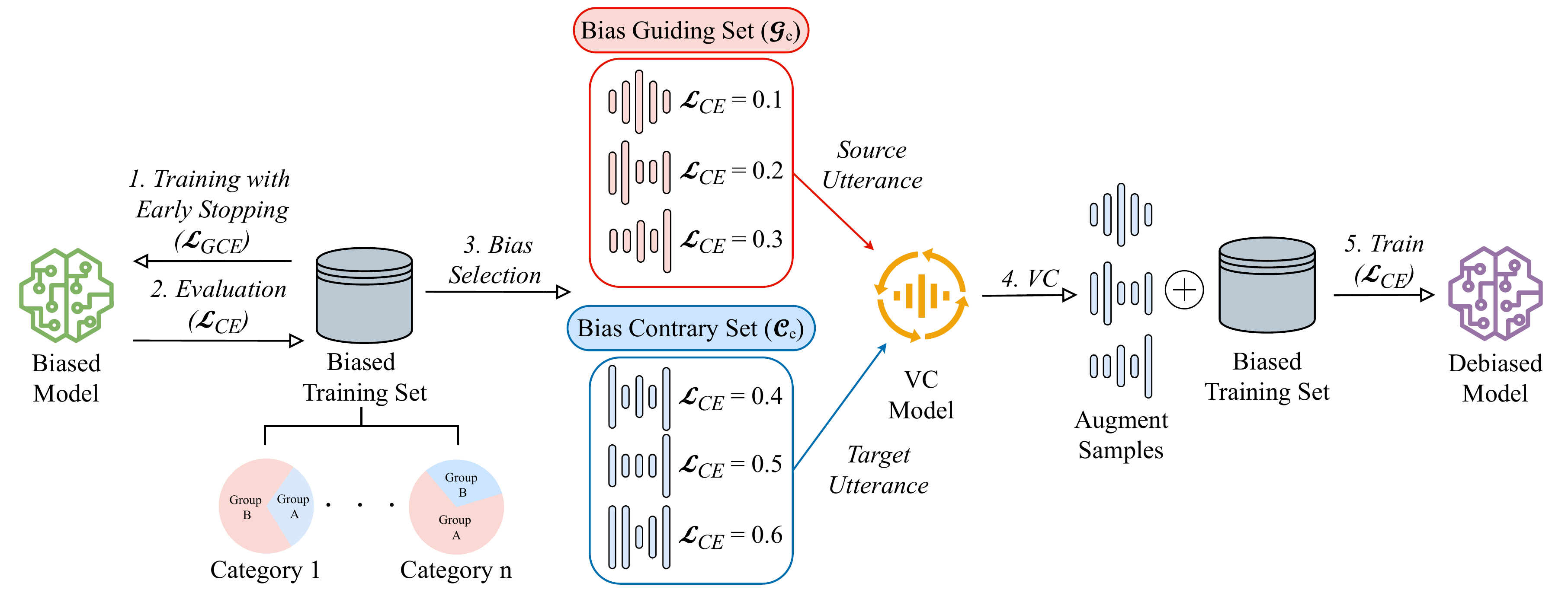}
  \vspace{-5mm}
  \caption{Overview of the proposed \textbf{CO-VADA}. $\mathcal{L}_{\mathrm{CE}}$ denotes the standard cross-entropy loss, and $\mathcal{L}_{\mathrm{GCE}}$ denotes the generalized cross-entropy loss used during early-stopped training.
  \emph{Category} refers to the emotion classes used as prediction targets, while \emph{Group} represents the speaker subgroup.
  After voice conversion, the resulting utterance retains the emotional content of the bias-guiding sample while adopting the speaker identity of the bias-contrary sample.}
  \label{fig:flow}
  \vspace{-3mm}
\end{figure*}

\vspace{-1mm}
\section{Introduction}
Speech emotion recognition (SER) systems are increasingly being applied in various fields, including customer service \cite{10388120, 10.1007/978-981-19-5868-7_14}, mental health monitoring \cite{10152117}, and human-computer interaction \cite{7172781}.
However, their deployment in real-world scenarios has raised growing concerns about their robustness and fairness \cite{10908859, chien24_interspeech}.
In particular, SER models are prone to spurious correlations between emotional labels and speaker characteristics (e.g., gender, age, race), which can emerge from unbalanced or biased training data \cite{lin24i_interspeech, slaughter-etal-2023-pre}.
For instance, if happy utterances are predominantly spoken by female speakers, the model may falsely associate vocal femininity with happiness.
These incorrect associations can lead to systematic misclassifications, especially for underrepresented groups, thereby introducing ethical and operational risks \cite{gorrostieta19_interspeech}.

However, to the best of our knowledge, very few studies have explored debiasing SER systems without relying on demographic information. Existing efforts in related speech tasks primarily focus on model-based interventions, such as adversarial training or sample reweighting \cite{jin22_interspeech, 8462663, kim23m_interspeech}, which often require architectural modifications and may lack generalizability.
In contrast, data-centric approaches (i.e., data augmentation or distribution balancing) tend to offer broader compatibility across models \cite{jin21_interspeech, zhang22n_interspeech}. 
Nevertheless, most of these methods still depend on explicit demographic annotations (e.g., gender, age, or speech disorder) to identify and mitigate bias. 
While these techniques have not been widely applied to SER, their general strategies hold promise for bias mitigation in this domain.
Unfortunately, in real-world applications, demographic metadata is often unavailable, incomplete, or unreliable. 
These limitations underscore the need for effective debiasing techniques that operate independently of demographic information.

To address the above-mentioned challenges, we propose \textbf{CO-VADA}, a \textit{Confidence-Oriented Voice Augmentation Debiasing Approach} for fair SER.
The core idea is to mitigate \textbf{speaker-related} bias without relying on demographic labels or architecture-specific modifications.
More specifically, \textbf{CO-VADA} utilizes model confidence, defined as the prediction certainty estimated by the loss during early training, along with speaker variation to guide targeted data augmentation.
This ad-hoc design provides a practical and generalizable solution for improving fairness in SER.

More precisely, we begin by training an early-stopped classifier on the original training data to estimate the prediction confidence for each sample. 
Within each emotion category, training samples are then grouped based on this confidence.
High-confidence samples are considered \textbf{bias-guiding} as they likely reflect dominant and potentially spurious patterns.
This assumption is based on the idea that early-stage models tend to rely on easily learnable and frequent patterns in the data, many of which may reflect spurious correlations or dataset bias \cite{LfF}.
In contrast, low-confidence samples are treated as \textbf{bias-contrary}, potentially capturing the traits of underrepresented speakers \cite{9710845}.
Augmented training examples are created by pairing samples from the previously defined bias-guiding and bias-contrary sets within each emotion category.
For each pair, we apply voice conversion to the bias-guiding sample, preserving its emotional content while adopting the speaker characteristics of the bias-contrary counterpart.
The converted samples are added to the training data to introduce the variation of the speaker and to encourage the model to focus on emotion-relevant features. 

Unlike prior debiasing methods in computer vision that rely on class activation maps or spatial structure to localize bias-inducing regions \cite{9710845}, CO-VADA tackles fundamentally different challenges in speech, where such spatial cues are unavailable.
Instead of modifying visual content, we operate in the audio domain by altering speaker identity to diversify training data. 
This targeted augmentation enables debiasing without demographic labels or architectural changes.

In summary, this work addresses speaker-related bias in SER and offers the following two contributions:
\begin{itemize}
    \item We present CO-VADA, a practical and extensible approach for debiasing SER systems.
    It enhances speaker diversity to improve fairness, without relying on demographic annotations or requiring any modifications to the model.

    \item We evaluate our method across multiple demographic attributes, including gender, race, and age.
    Extensive experiments on three benchmark datasets demonstrate its effectiveness in improving fairness while maintaining strong overall performance.

\end{itemize}

\vspace{-1mm}
\section{Related Work}
\vspace{-1mm}
\subsection{Bias in Speech Systems}
Previous work has documented a range of biases in speech systems, most of which are speaker biases, arising from individual voice characteristics or demographic attributes rather than environmental or content factors.
For instance, gender bias manifests as higher word error rates (WER) in Automatic Speech Recognition (ASR) \cite{tatman-2017-gender, electronics11101594}, higher macro-F1 score for female speakers in SER \cite{lin24i_interspeech}, and social stereotypical associations of gender in speech large language models \cite{10832259}. 
Also, accent and dialect bias lead to poorer recognition for non-native or regional accents, including Scottish English \cite{tatman-2017-gender}, non-American accents \cite{zhang22n_interspeech, 10887767}, and African American English \cite{10.1093/applin/amac066}. 
Age bias appears with children’s and elderly voices that produce more recognition errors than adult voices \cite{vipperla2010ageing, attia2024kid}, and a spurious correlation in the embedding space \cite{lin24b_interspeech}. Disability bias results in significantly worse performance on speech from speakers with disorders such as dysarthria \cite{10.1121/1.4967208, kim23m_interspeech}. 
Since these disparities originate from who speaks, regarding their speech patterns or demographic group, we also consider speaker-centric bias in our work.

\vspace{-1mm}
\subsection{Debiasing Methods with Demographic Information}
Recent debiasing methods for speech-input tasks rely on a variety of techniques, including counterfactual data augmentation through voice conversion \cite{9610166}, cross-lingual adaptation combined with SpecAugment and speed perturbation \cite{zhang22n_interspeech, 10389756}, and subgroup-aware sampling guided by divergence metrics \cite{10446326, dheram22_interspeech}. 
Besides, other approaches include group-adapted fusion networks for fairer speaker verification (SV) \cite{9747384}, adversarial training \cite{kim2025domain, 10889760}, gender neutralization \cite{electronics11101594}, and contrastive learning to enhance subgroup-invariant representations in intent classification \cite{koudounas24b_interspeech}. 
While these techniques have successfully reduced demographic performance gaps in ASR, SV, and Spoken Language Understanding tasks, they rely on predefined demographic labels, which limit their applicability to unannotated or intersectional biases.

\vspace{-1mm}
\subsection{Debiasing Methods without Demographic Information}
Recent advancements in computer vision and natural language processing have addressed the challenge of debiasing without the use of demographic annotations, instead leveraging inherent training signals or auxiliary objectives. 
For instance, LVR \cite{LVR} implements class-wise low-variance regularization to compress embeddings and diminish spurious features. 
LfF \cite{LfF} employs a two-step approach where a deliberately “prejudiced” network is first trained to enhance spurious correlations, followed by the training of a second model that focuses on correcting the biases by learning from the first model's errors. 
Building on this foundation, DisEnt \cite{DisEnt} introduces disentangled feature augmentation, which synthesizes diverse “bias-conflicting” samples. 
In this method, two encoders separate intrinsic features from bias correlations identified in LfF within the latent space, allowing for the swapping of these features to create enriched counterexamples. 

Additionally, SiH \cite{vandenhirtz2023signal} utilizes a Variational Autoencoder (VAE) combined with focal-loss reweighting to develop both biased and unbiased classifiers, jointly down-weighting spurious patterns. 
BLIND \cite{Blind} trains a “success detector” that predicts when the main model’s decisions hinge on simplistic, shortcut features, subsequently down-weighting these samples through a debiased focal loss \cite{karimi-mahabadi-etal-2020-end}, thereby dynamically alleviating bias using only task labels.

In the realm of speech processing, previous studies have exploited embedding clustering for the unsupervised investigation of biased associations. 
Implicit Demography Inference \cite{lin2025mitigatingsubgroupdisparitiesmultilabel} utilizes k-means \cite{kmeans} to analyze ECAPA-TDNN \cite{desplanques20_interspeech} speaker embeddings, uncovering latent speaker groups and training SER models with group-aware objectives, which significantly reduces subgroup disparities. 
This approach is also applied in ASR \cite{veliche2023improving} and in ensuring individual fairness in dimensional SER (e.g., arousal/valence predictions) \cite{chou24_interspeech}. 

Unlike previous work \cite{lin2025mitigatingsubgroupdisparitiesmultilabel}, which clusters training data based on speaker-related embeddings, the proposed CO-VADA achieves better performance by directly identifying bias-guiding and bias-contrary samples.
Instead of simply adjusting sample weights or appending cluster labels, we apply voice conversion to generate diverse audio samples. 
Each converted utterance preserves the original emotional content while adopting the acoustic characteristics of an underrepresented speaker, resulting in more effective debiasing.

\vspace{-1mm}
\section{Methodology}

This section details five steps of the proposed CO-VADA, shown in Fig.~\ref{fig:flow}.

\vspace{-1mm}
\subsection{Sample Separation by Confidence (\textbf{Step 1-3})}

We started by training a bias selection classifier using the original dataset \( \mathcal{D}_{\text{orig}} \) (\textbf{Step~1} in Fig.~\ref{fig:flow}).
The training employed a \textbf{class-balanced generalized cross-entropy loss} \cite{10.1145/3583780.3615184, pmlr-v139-liu21f}.
We implemented early stopping once the model reached moderate performance.
At this point, the model was more likely to rely on dominant and easily learnable patterns \cite{LfF}, which may include spurious correlations between speaker characteristics and emotion labels.

To analyze this potential bias, we define the emotion-specific subsets of the data.
Let $\mathrm{\textit{E}}$ be the set of all emotion labels in the dataset.
For each emotion category \( e \in E \), we extract the label subset:
\begin{equation}
\label{equ:emotion_label}
\mathcal{D}_e = \{ x \mid x \in \mathcal{D}_{\text{orig}} \land  y_e = 1 \},
\end{equation}
where \( y_e \in \{0, 1\} \) is the binary ground-truth label for emotion \( e \) associated with sample \( x \).  
That is, a sample \( x \) belongs to \( \mathcal{D}_e \) if it is expressing emotion \( e \).

Each sample \( x \in \mathcal{D}_e \) is evaluated using the \textbf{cross-entropy loss} ($\mathcal{L}_{\mathrm{CE}}$) between the model prediction and the true label, which serves as a proxy for prediction uncertainty (\textbf{Step~2} in Fig.~\ref{fig:flow}). Lower $\mathcal{L}_{\mathrm{CE}}$ indicates higher model confidence.
We ranked the samples in \( \mathcal{D}_e \) by their $\mathcal{L}_{\mathrm{CE}}$ values. The more confidently predicted ones (with lower $\mathcal{L}_{\mathrm{CE}}$) form the \textbf{bias-guiding set} \( \mathcal{G}_e \), assumed to reflect dominant patterns. In contrast, those with higher loss values constitute the \textbf{bias-contrary set} \( \mathcal{C}_e \), potentially representing underrepresented speaker traits or atypical characteristics (\textbf{Step~3} in Fig.~\ref{fig:flow}).
Due to the multi-label nature of the dataset, a single sample \( x \) may belong to multiple subsets \( \mathcal{D}_e \), and be categorized differently across emotion labels.
Since the proposed approach only uses $\mathcal{L}_{\mathrm{CE}}$ to select bias-guiding and bias-contrary sets, it does not require knowledge of the demographics of the training examples.

\vspace{-1mm}
\subsection{Voice Conversion and Augmentation (\textbf{Step 4-5})}

For each emotion label \( e \in E \), we randomly sampled \( K \) utterance pairs, with one utterance drawn from the bias-guiding set \( \mathcal{G}_e \) and the other from the bias-contrary set \( \mathcal{C}_e \).  
In each pair, the utterance from \( \mathcal{G}_e \) served as the \textbf{source utterance}, preserving its emotional content, while the utterance from \( \mathcal{C}_e \) acted as the \textbf{target utterance}, providing the speaker characteristics to be transferred through voice conversion (\textbf{Step~4} in Fig.~\ref{fig:flow}).  
The converted sample inherited the emotion label from the source and incorporated the speaker traits of the target, thereby introducing speaker variability to support bias mitigation.

All converted samples were added to the original training set \( \mathcal{D}_{\text{orig}} \) to form the augmented set \( \mathcal{D}_{\text{aug}} \).
We used \( \mathcal{D}_{\text{aug}} \) to train a new classifier. 
This model learned from a wider range of diverse instances, which had less bias for each emotion.
As a result, the model was encouraged to rely on emotion-relevant features rather than on spurious cues such as speaker characteristics (\textbf{Step~5} in Fig.\ref{fig:flow}).

\vspace{-1mm}
\section{Evaluation}

To evaluate both recognition performance and fairness in SER, we utilized three key metrics: macro-F1 score (referred to as \textbf{Macro F1}), \textbf{True Positive Rate Gap} ($\mathrm{TPR}_{\text{gap}}$) \cite{han-etal-2022-balancing, chen-etal-2024-addressing}, and \textbf{Demographic Parity Gap} ($\mathrm{DP}_{\text{gap}}$) \cite{chen-etal-2024-addressing, 10446564}. 

\vspace{-1mm}
\subsection{Macro-F1 Score}
Macro F1 measures the overall classification quality by treating all emotion classes equally, which is crucial in SER due to issues with class imbalance. 

\vspace{-1mm}
\subsection{True Positive Rate Gap}

To assess fairness in classification performance across demographic groups, we compute \( \mathrm{TPR}_{\text{gap}} \) \cite{NIPS2016_9d268236}.  
Let \( \mathcal{Z} \) denote the set of demographic groups, and \( E \) the set of emotion classes.
For each class \( e \in E \) and group \( z \in \mathcal{Z} \), the true positive rate is defined as \( \mathrm{TPR}_{e}^{(z)} \).  

Let \( P \) be the total number of unordered group pairs across all emotion classes. The overall \( \mathrm{TPR}_{\text{gap}} \) is computed as:
\begin{equation}
\label{equ:tpr_gap}
\mathrm{TPR}_{\text{gap}} = \sqrt{ \frac{1}{P} \sum_{e \in E} \sum_{z_i \ne z_j} 
\left| \mathrm{TPR}_{e}^{(z_i)} - \mathrm{TPR}_{e}^{(z_j)} \right|^2 },
\end{equation}
where \( z_i, z_j \in \mathcal{Z} \) denote demographic groups and \( z_i \ne z_j \).  
A smaller \( \mathrm{TPR}_{\text{gap}} \) indicates more consistent sensitivity across groups, reflecting greater fairness in emotion recognition.

\vspace{-1mm}
\subsection{Demographic Parity Gap}

We evaluate output fairness using $\mathrm{DP}_{\text{gap}}$.  
Let \( N \) be the total number of samples. For each class \( e \in E \), let \( \hat{y}_{le} \in \{0, 1\} \) denote the predicted label for sample \( l \) in class \( e \).
The global positive prediction rate for class \( e \) is defined as follows:
\begin{equation}
\label{equ:y_global}
\hat{y}_e^{\mathrm{global}} = \frac{1}{N} \sum_{l=1}^{N} \hat{y}_{le}.
\end{equation}
Although normalizing by class-specific sample counts may seem reasonable, it’s important to note that in a multi-label setting, every sample can be associated with all classes.
Therefore, we consistently use the total number of samples \( N \) for all classes.

Let \( N_z \) be the number of samples belonging to group \( z \).  
The group prediction rate for group \( z \) is defined as follows:
\begin{equation}
\label{equ:y_group}
\hat{y}_e^{(z)} = \frac{1}{N_z} \sum_{l=1}^{N_z} \hat{y}_{le},
\end{equation}
where the summation is taken over all samples \( x_l \) whose demographic group is \( z \).

Finally, the overall \( \mathrm{DP}_{\text{gap}} \) is computed as:
\begin{equation}
\label{equ:dp_gap}
\mathrm{DP}_{\text{gap}} = \sqrt{ \frac{1}{|E|} \sum_{e \in E} \max_{z \in \mathcal{Z}} \left| \hat{y}_e^{(z)} - \hat{y}_e^{\mathrm{global}} \right|^2 }.
\end{equation}
Lower \( \mathrm{DP}_{\text{gap}} \) values indicate more uniform prediction rates across different demographic groups, leading to better output fairness.

\input{result}

\begin{table}[t]
    \renewcommand{\arraystretch}{1.2}
    \centering
    \caption{Sample counts for each emotion and gender (Male/Female) in the CREMA-D dataset across train, development, and test sets (shown for Fold 1 defined in the EMO-SUPERB\cite{Wu_2024}).}
    \vspace{-3mm}
    \label{tab:cremad_distribution}
    \fontsize{7}{9}\selectfont
    \begin{tabular}{lc@{/}c c@{/}c c@{/}c}
    \toprule
    \textbf{Emotion} & \multicolumn{2}{c}{\textbf{Train (M/F)}} & \multicolumn{2}{c}{\textbf{Dev (M/F)}} & \multicolumn{2}{c}{\textbf{Test (M/F)}} \\
    \midrule
    Angry   & 240 & 12   & 93 & 4    & 58 & 102  \\
    Disgust & 86  & 4    & 34 & 1    & 18 & 41   \\
    Neutral & 1080 & 54  & 384 & 19  & 250 & 263 \\
    Fear    & 6   & 127  & 1  & 27   & 20 & 55   \\
    Happy   & 5   & 102  & 1  & 24   & 15 & 35   \\
    Sad     & 2   & 58   & 2  & 41   & 7  & 35   \\
    \bottomrule
    \end{tabular}
    \vspace{-4mm}
\end{table}

\vspace{-1mm}
\section{Experimental Setup}

\vspace{-1mm}
\subsection{Model Setup}

In our experiments, we used WavLM-base+~\footnote{https://github.com/microsoft/unilm/tree/master/wavlm} \cite{9814838} as the feature extractor. 
We applied a learnable weighted sum over all feature layers to obtain the final representation \cite{yang21c_interspeech}, then passed it on to a simple two-layer linear classifier.
We trained all models using AdamW \cite{loshchilov2019decoupledweightdecayregularization} optimizer with a learning rate of 1e-4 and a batch size of 32.

\vspace{-1mm}
\subsection{Datasets}
We evaluated our framework using three well-known SER datasets: \textbf{CREMA-D} \cite{6849440}, \textbf{MSP-Podcast} \cite{8003425}, and \textbf{MSP-IMPROV} \cite{7374697}. 
All three datasets employed a multi-label annotation scheme and cover a variety of emotional categories, as noted in \cite{Wu_2024, Chou_2025}. 
In our study, we used only the primary emotions (single choice of options for each annotator) from the MSP-IMPROV and MSP-Podcast datasets.
We defined three different emotion classification schemes: a 4-class classification for MSP-IMPROV, a 6-class classification for CREMA-D, and an 8-class classification for MSP-Podcast. 
Further details can be found in \cite{Wu_2024}.

To simulate biased learning conditions, we manually introduced gender-emotional imbalance in the training and development sets. 
Specifically, for each emotion category, we skewed the gender ratio to approximately 1:20 \cite{lin2025mitigatingsubgroupdisparitiesmultilabel}. 
Please note that the information about bias is not used in the proposed approach.
Table~\ref{tab:cremad_distribution} provides an example of this imbalance configuration in the CREMA-D dataset. 
The test sets were kept unchanged, following the original splits, and serve as unbiased references for evaluation.

While demographic labels are not used during model training, we utilized available metadata during evaluation to compute fairness metrics. 
The three datasets differ in their demographic coverage and the scope of fairness analysis. 
The CREMA-D included annotations for gender, age, and race, enabling evaluation on multiple demographic axes. 
The MSP-Podcast and MSP-IMPROV only provided gender metadata; thus, fairness evaluation on these datasets was restricted to gender-based analysis.
In addition, CREMA-D was partitioned into five folds and MSP-IMPROV into six folds, respectively, following the cross-validation setting in EMO-SUPERB \cite{Wu_2024}.

\vspace{-1mm}
\subsection{Voice Conversion Models}

VC was used in our framework to generate speaker-varied audio samples for data augmentation. 
We primarily adopted \textbf{FreeVC} \cite{10095191} for all main experiments, due to its high-quality, text-free, one-shot conversion capability using WavLM features. 

To evaluate the generality of our framework across various VC implementations, we included two additional models in our ablation studies: Diff-HierVC \cite{choi23d_interspeech}, which is a diffusion-based model featuring a hierarchical architecture, and kNN-VC \cite{baas23_interspeech}, a nonparametric method that performs frame-level matching without training.
We directly applied all VC models in their pretrained form, without any fine-tuning or model-specific optimization.

\vspace{-1mm}
\subsection{Bias Partitioning and Augmentation Strategy}

To identify samples that reflect biases at the dataset level, we first trained an initial emotion classifier with early stopping. 
The main purpose is to capture the model's early learning dynamics, a phase in which it tends to rely on spurious correlations. 
Early stopping was triggered once the Macro F1 exceeded a predefined threshold of 0.5 for the CREMA-D and MSP-IMPROV, and 0.3 for the MSP-Podcast. 
The lower threshold for MSP-Podcast reflected the particular difficulty in achieving high performance in SER using this dataset.
In our experiments, model performance on the MSP-Podcast rarely exceeds a Macro F1 of 0.5 during training.

We adopted a consistent strategy for pairing and sample generation.  
Within each emotion category, training samples were ranked in ascending order by cross-entropy loss.  
The lower half of the ranked samples was designated as the bias-guiding set, while the upper half formed the bias-contrary set.  
We then created random one-to-one pairings between the two sets.  
For each emotion category, we generated a number of augmented samples equal to the count of original samples in that category.

Due to the multi-label nature of the dataset, some utterances may participate in multiple emotion categories. 
To obtain binary indicators from soft labels, we assigned an emotion as present if its label score exceeds a threshold of $\frac{1}{\left | E \right |}$, following the previous works \cite{Wu_2024,Chou_2025}.  
As a result, the total number of converted samples can exceed the size of the original training set.
All converted audio was merged with the original training set and used in training without distinction.

\vspace{-1mm}
\subsection{Baselines and Selection Criteria}

We compared CO-VADA with six representative de-biasing baselines: \textbf{LVR} \cite{LVR}, \textbf{SiH} \cite{vandenhirtz2023signal}, \textbf{BLIND} \cite{Blind}, \textbf{IDI-RW} \cite{lin2025mitigatingsubgroupdisparitiesmultilabel}, \textbf{DisEnt} \cite{DisEnt}, and \textbf{LfF} \cite{LfF}. 
These methods cover a range of debiasing strategies, including reweighting, disentangled representation learning, adversarial training, and selective sampling. 
We focused on approaches that do not rely on demographic labels, ensuring consistency with our training setting.
To allow a fair comparison, all methods were evaluated under the same experimental configuration. 
The ERM was the original baseline without any debiasing method, following \cite{lin2025mitigatingsubgroupdisparitiesmultilabel}.

To ensure a fair and comprehensive comparison, we evaluated each method under five hyperparameter settings, avoiding selective tuning that could bias the comparison.
For each method, we selected one key tunable parameter and tested five different values.

\vspace{-1mm}
\section{Experimental Results}

The proposed \textbf{CO-VADA} achieved the best balance between classification performance and fairness across gender, race, and age based on results from the CREMA-D dataset. 
Fig.~\ref{fig:cremad-result} illustrates these results in performance-fairness scatter plots. 
It appeared consistently in the \textbf{lower right region} of the F1–$\mathrm{TPR}_{\text{gap}}$ and F1–$\mathrm{DP}_{\text{gap}}$ scatter plots, indicating strong predictive accuracy with low subgroup disparities.
Among the baselines, SiH, DisEnt, and LfF reached similar Macro F1 but exhibited substantially larger fairness gaps, suggesting that these methods did not effectively mitigate bias. 
Moreover, LVR performed comparably with ours in fairness, particularly for gender, but suffered from a noticeable drop in Macro F1. 
IDI-RW approached our fairness levels, yet its classification performance remained consistently lower. 
ERM, which does not employ a debiasing strategy, yielded moderate accuracy but poor fairness. 

Regarding the results on the MSP-Podcast, our method achieved the lowest $\mathrm{TPR}_{\text{gap}}$ among all baselines. 
Fig.~\ref{fig:podcast-improv-fairness} shows the corresponding plots for gender fairness. 
These findings demonstrated that our method achieved the most consistent recognition of emotions between gender subgroups, which is particularly important in fairness-sensitive applications. 
While our Macro F1 was slightly lower than LVR's, the difference was negligible.
LVR and our method demonstrated comparable $\mathrm{TPR}_{\text{gap}}$ and $\mathrm{DP}_{\text{gap}}$ values, indicating similar fairness performance across demographic groups.
Besides, SiH achieved a marginal improvement in Macro F1, but exhibited substantial fairness gaps. 
LfF also did not match our method in either metric. 

In terms of results on the MSP-IMPROV, IDI-RW achieved the best overall result in both fairness and SER performance. 
Fig.~\ref{fig:podcast-improv-fairness} shows the corresponding graphs. 
It showed the lowest $\mathrm{TPR}_{\text{gap}}$ and $\mathrm{DP}_{\text{gap}}$, along with the highest Macro F1 among all methods. 
However, IDI-RW performed poorly on the CREMA-D and MSP-Podcast, suggesting that its effectiveness may be limited to certain data conditions. 
Following IDI-RW, our CO-VADA offered the next best trade-off, achieving consistently low $\mathrm{TPR}_{\text{gap}}$ and $\mathrm{DP}_{\text{gap}}$ along with competitive Macro F1. 
Other baselines, such as LVR, SiH, DisEnt, and LfF, performed worse in terms of both fairness and accuracy. 
These methods tended to cluster toward the upper-middle region of the plot.

Across all three datasets, \textbf{the CO-VADA consistently achieved substantial trade-offs between fairness and SER performance.} 
It maintained competitive or superior Macro F1 while substantially reducing both $\mathrm{TPR}_{\text{gap}}$ and $\mathrm{DP}_{\text{gap}}$, particularly under conditions of group-level imbalance. 
In contrast, other debiasing methods often sacrificed fairness for accuracy or vice versa. 
The stable performance of our method highlighted its robustness to data variation and confirmed its generalizability in fairness-aware emotion recognition.

\begin{table}[t]
    \renewcommand{\arraystretch}{1.2}
    \centering
    \caption{Performance under different macro F1 thresholds used for early stopping.
    Each row shows the full debiasing pipeline result using a different threshold.
    $\uparrow$ indicates higher is better, while $\downarrow$ indicates lower is better.}
    \vspace{-2mm}
    \fontsize{7}{9}\selectfont
    \begin{tabular}{cccc}
        \toprule
        \textbf{Threshold} & \textbf{Macro F1} $\uparrow$ & \textbf{$\boldsymbol{\mathrm{TPR}_{\text{gap}}}$} $\downarrow$ & \textbf{$\boldsymbol{\mathrm{DP}_{\text{gap}}}$} $\downarrow$ \\\midrule
        0.3 & 0.653\greensub{+0.31\%} & \textbf{0.231}\greensub{\textbf{-16.61\%}} & 0.088\greensub{-14.56\%} \\
        0.4 & \textbf{0.658}\greensub{\textbf{+1.07\%}} & 0.245\greensub{-11.55\%} & 0.089\greensub{-14.56\%} \\
        0.5 & 0.653\greensub{+0.31\%} & 0.233\greensub{-15.88\%} & \textbf{0.082}\greensub{\textbf{-20.39\%}} \\
        0.6 & 0.655\greensub{+0.61\%} & 0.235\greensub{-15.16\%} & 0.090\greensub{-12.62\%} \\
        0.7 & 0.653\greensub{+0.31\%} & 0.233\greensub{-15.88\%} & 0.087\greensub{-15.53\%} \\
        \bottomrule
    \end{tabular}
    \label{tab:threshold_raw}
    \vspace{-3mm}
\end{table}

\vspace{-1mm}
\section{Ablation Study}

We conducted a series of ablation experiments to examine how different design choices affect the SER performance and fairness of our debiasing framework. 
All experiments were performed on the CREMA-D dataset, with a focus on gender-based fairness metrics.
The training configuration remained identical to that described in the main experimental setup.

\vspace{-1mm}
\subsection{Sensitivity of Model Performance to Early Stopping Thresholds}

To evaluate the effect of early stopping on bias mitigation, we varied the Macro F1 threshold used to trigger early termination during training on the CREMA-D.
The thresholds ranged from 0.3 to 0.7.
The results in Table\ref{tab:threshold_raw} show the performance of CO-VADA under each threshold, \textbf{with subscripts indicating the relative improvements over the baseline ERM (green boxes denote improvements over ERM; red boxes denote degradations)}.
Macro F1 consistently improved across all thresholds, and fairness metrics also showed reductions relative to ERM, though the extent of improvement varied slightly across thresholds. 
These findings indicate that the specific choice of the early stopping threshold has a limited effect on overall performance and fairness. 
CO-VADA remains effective in mitigating bias even when the threshold is coarsely selected.

\vspace{-1mm}
\subsection{Impact of Bias Proportion in Training Data}

We evaluated how different proportions of bias-guiding and bias-contrary samples influence model performance and fairness. 
As shown in Table~\ref{tab:ratio}, all configurations yielded comparable Macro F1, suggesting that classification performance was unaffected by the sampling ratio. 
However, slight but consistent differences appeared in $\mathrm{TPR}_{\text{gap}}$ and $\mathrm{DP}_{\text{gap}}$. 
This pattern suggested that fairness was more sensitive than accuracy to the composition of augmented samples.

\begin{table}[t]
    \centering
    \renewcommand{\arraystretch}{1.2}
    \caption{Performance under different \textbf{bias-contrary:unused:bias-guiding} ratios. 
    Subscripts indicate relative improvements over the ERM baseline.}
    \vspace{-2mm}
    \fontsize{7}{9}\selectfont
    \begin{tabular}{cccc}
        \toprule
        \textbf{Ratio} & \textbf{Macro F1} $\uparrow$ & \textbf{$\boldsymbol{\mathrm{TPR}_{\text{gap}}}$} $\downarrow$ & \textbf{$\boldsymbol{\mathrm{DP}_{\text{gap}}}$} $\downarrow$ \\
        \midrule
        3 : 4 : 3 & 0.654\textsubscript{\greensub{+0.46\%}} & 0.230\textsubscript{\greensub{-16.96\%}} & 0.087\textsubscript{\greensub{-15.53\%}} \\
        4 : 2 : 4 & \textbf{0.655}\textsubscript{\textbf{\greensub{+0.61\%}}} & 0.247\textsubscript{\greensub{-10.83\%}} & 0.092\textsubscript{\greensub{-10.68\%}} \\
        5 : 0 : 5 & 0.653\textsubscript{\greensub{+0.31\%}} & 0.233\textsubscript{\greensub{-15.88\%}} & \textbf{0.082}\textsubscript{\textbf{\greensub{-20.39\%}}} \\
        4 : 0 : 6 & 0.653\textsubscript{\greensub{+0.31\%}} & 0.240\textsubscript{\greensub{-13.36\%}} & 0.088\textsubscript{\greensub{-14.56\%}} \\
        3 : 0 : 7 & 0.649\textsubscript{\greensub{-0.31\%}} & \textbf{0.227}\textsubscript{\textbf{\greensub{-18.05\%}}} & 0.083\textsubscript{\greensub{-19.42\%}} \\
        \bottomrule
    \end{tabular}
    \label{tab:ratio}
    \vspace{-2mm}
\end{table}

\vspace{-1mm}
\subsection{Effectiveness of Different Voice Conversion Models}

We evaluated the effectiveness of CO-VADA using different VC models.
As shown in Table \ref{tab:vc_models}, FreeVC achieved the best overall balance in relative improvement over the ERM baseline. 
It yielded the highest F1 gain while substantially reducing both TPR\textsubscript{gap} and DP\textsubscript{gap}.
Diff-HierVC achieved the largest reduction in fairness gaps, though with a slight drop in F1. 
kNN-VC also improved fairness compared to ERM, but showed a relative decrease in classification performance, possibly due to its nonparametric nature.

While the degree of improvement varied depending on the generation quality of each VC model, all models consistently enhanced fairness. This confirms that the effectiveness of our debiasing approach stems from the overall design of CO-VADA, rather than reliance on any particular VC model.

\begin{table}[t]
    \centering
    \renewcommand{\arraystretch}{1.2}
    \caption{SER performance and fairness using different VC models. 
    Subscripts indicate relative improvement over the ERM baseline.}
    \vspace{-2mm}
    \fontsize{7}{9}\selectfont
    \begin{tabular}{lccc}
    \toprule
    \textbf{VC Model} & \textbf{Macro F1} $\uparrow$ & \textbf{$\boldsymbol{\mathrm{TPR}_{\text{gap}}}$} $\downarrow$ & \textbf{$\boldsymbol{\mathrm{DP}_{\text{gap}}}$} $\downarrow$ \\
    \midrule
    kNN-VC\cite{baas23_interspeech}       & 0.642\textsubscript{\redsub{-1.38\%}} & 0.244\textsubscript{\greensub{-12.23\%}} & 0.097\textsubscript{\greensub{-5.83\%}} \\
    Diff-HierVC\cite{choi23d_interspeech} & 0.648\textsubscript{\redsub{-0.46\%}} & \textbf{0.223}\textsubscript{\textbf{\greensub{-19.78\%}}} & \textbf{0.081}\textsubscript{\textbf{\greensub{-21.36\%}}} \\
    FreeVC\cite{10095191}                 & \textbf{0.653}\textsubscript{\textbf{\greensub{+0.31\%}}} & 0.233\textsubscript{\greensub{-16.19\%}} & 0.082\textsubscript{\greensub{-20.39\%}} \\
    \bottomrule
    \end{tabular}
    \label{tab:vc_models}
    \vspace{-3mm}
\end{table}

\begin{table}[t]
    \centering
    \renewcommand{\arraystretch}{1.2}
    \caption{SER performance and fairness under different combinations of BS and VC. 
    Subscripts indicate relative improvements over the ERM baseline. \ding{51} indicates the component is included.}
    \label{tab:bs_vc_ablation}
    \vspace{-2mm}
    \fontsize{7}{9}\selectfont
    \begin{tabular}{c c c c c c}
        \toprule
        \# & \textbf{BS} & \textbf{VC} & \textbf{Macro F1} $\uparrow$ & \textbf{$\boldsymbol{\mathrm{TPR}_{\text{gap}}}$} $\downarrow$ & \textbf{$\boldsymbol{\mathrm{DP}_{\text{gap}}}$} $\downarrow$ \\
        \midrule
        1 & \ding{51} & \ding{53} & \textbf{0.659}\textsubscript{\textbf{\greensub{+1.23\%}}} & 0.259\textsubscript{\greensub{-6.83\%}}  & 0.097\textsubscript{\greensub{-5.83\%}} \\
        2 & \ding{53} & \ding{51} & 0.652\textsubscript{\greensub{+0.15\%}} & 0.240\textsubscript{\greensub{-13.67\%}} & 0.088\textsubscript{\greensub{-14.56\%}} \\
        3 & \ding{51} & \ding{51} & 0.653\textsubscript{\greensub{+0.31\%}} & \textbf{0.233}\textsubscript{\textbf{\greensub{-16.19\%}}} & \textbf{0.082}\textsubscript{\textbf{\greensub{-20.39\%}}} \\
        \bottomrule
    \end{tabular}
    \vspace{-2mm}
\end{table}

\subsection{Effect of Bias Selection and Voice Conversion}

We compared three configurations that vary in the use of bias selection (\textbf{BS}) and voice conversion (\textbf{VC}) augmentation.
Table \ref{tab:bs_vc_ablation} reports their relative improvement over the ERM baseline, which does not apply BS or VC.

\begin{itemize}
    \item Configuration\#1 included BS but not VC. In this setting, bias-contrary samples were assigned higher weights during training, but no data were augmented. 
    \item Configuration\#2 included VC but not BS. VC was applied to randomly selected pairs of utterances without considering their bias status. 
    \item Configuration\#3, which represents our complete method, applied both BS and VC by selecting bias-guiding and bias-contrary samples and generating converted examples accordingly.
\end{itemize}

As shown in Table \ref{tab:bs_vc_ablation}, our method (\#3) achieved the best overall improvement compared to ERM, striking the best balance between classification performance and fairness.
It produced the largest reductions in $\mathrm{TPR}_{\text{gap}}$ and $\mathrm{DP}_{\text{gap}}$, along with competitive Macro F1. 
Configuration\#1 improved the accuracy slightly but offered a limited fairness improvement. 
Configuration\#2 yielded more fairness improvement than \#1, but its overall results remained inferior to ours.

\vspace{-1mm}
\subsection{SER Performance and Fairness Across Emotion Categories}

We further evaluated the SER performance and fairness of CO-VADA across different emotion categories.
As shown in Table \ref{tab:emotion}, we report the original results of CO-VADA for each emotion class, with subscripts indicating the relative changes compared to the corresponding ERM baseline for that class. That is, each comparison uses the ERM result specific to the same emotion category, rather than the overall ERM average reported earlier. 

The proposed CO-VADA consistently reduced gender-based disparities across emotions, while maintaining competitive classification performance in most categories. These results demonstrate the robustness of CO-VADA in handling diverse emotional expressions.

\begin{table}[t]
\centering
\caption{SER performance and gender fairness of CO-VADA across emotion categories. 
Subscripts indicate relative changes compared to the ERM baseline.}
\vspace{-2mm}
\fontsize{7}{9}\selectfont
\begin{tabular}{lccc}
\toprule
\textbf{Emotion} & \textbf{F1 $\uparrow$} & \textbf{TPR\textsubscript{gap} $\downarrow$} & \textbf{DP\textsubscript{gap} $\downarrow$} \\
\midrule
Angry   & 0.534\textsubscript{\redsub{-1.48\%}} & 0.270\textsubscript{\greensub{-2.53\%}}  & 0.104\textsubscript{\redsub{+7.22\%}}   \\
Happy   & 0.540\textsubscript{\greensub{+0.93\%}} & 0.374\textsubscript{\greensub{-1.32\%}}  & 0.118\textsubscript{\greensub{-11.94\%}} \\
Sad     & 0.536\textsubscript{\redsub{-0.92\%}} & 0.229\textsubscript{\greensub{-4.58\%}}  & 0.090\textsubscript{\greensub{-21.05\%}} \\
Fear    & 0.594\textsubscript{\greensub{0.00\%}}  & 0.229\textsubscript{\greensub{-17.92\%}} & 0.103\textsubscript{\greensub{-20.77\%}} \\
Neutral & 0.606\textsubscript{\redsub{-0.82\%}} & 0.247\textsubscript{\greensub{-15.71\%}} & 0.080\textsubscript{\greensub{-23.81\%}} \\
Disgust & 0.576\textsubscript{\redsub{-0.17\%}} & 0.224\textsubscript{\greensub{-17.58\%}} & 0.081\textsubscript{\greensub{-18.18\%}} \\
\bottomrule
\end{tabular}
\label{tab:emotion}
\vspace{-3mm}
\end{table}

\section{Conclusion, Limitations, and Future Work}
Experiments on multiple benchmark datasets showed that our \textbf{CO-VADA} consistently improves fairness while maintaining or improving classification performance. 
The framework is compatible with different voice conversion models and can be seamlessly integrated into standard training pipelines, making it effective and broadly applicable to fairness-aware SER tasks.

A key limitation of our approach is the reliance on prediction confidence to separate bias-guiding and bias-contrary samples. 
While effective in practice, this heuristic may conflate underrepresented but easy samples with overrepresented but difficult ones.

Moreover, while our method is not limited to gender bias, due to the limitations of the corpus and its labels, we evaluated the proposed approach only on a dataset that contains gender bias.
Future work may explore other sources of bias, such as semantic \cite{10832317}, linguistic \cite{10887615}, or stylistic variation \cite{9747897}.
Besides, investigate alternative debiasing strategies that operate in the latent space or directly manipulate embeddings potentially enabling more efficient and flexible mitigation without relying on audio synthesis.

\bibliographystyle{IEEEtran}
\bibliography{IEEEref}

\end{document}